%% file: FRM.tex
\documentclass[apj,numberedappendix,appendixfloats,tighten,twocolumn]{emulateapj}
\input{Packages.tex}
\input{Definitions.tex}

\begin{document}


\title{\MakeUppercase{What can we really learn about Magnetic Fields in Galaxy Clusters from Faraday Rotation observations?}}

\author{Gilad Rave\altaffilmark{1}, Doron Kushnir\altaffilmark{2}, and Eli Waxman\altaffilmark{1}}
\altaffiltext{1}{Department of Particle Physics and Astrophysics, Weizmann Institute of Science, Rehovot 76100, Israel}
\altaffiltext{2}{Institute for Advanced Study, Einstein Drive, Princeton, New Jersey, 08540, USA}

\begin{abstract}

We construct a simple and robust approach for deriving constraints on magnetic fields in galaxy clusters from rotation measure (RM) maps. Relaxing the commonly used assumptions of a correlation between the magnetic field strength and the plasma density and of a power-law (in wave number) magnetic field power spectrum, and using an efficient numerical analysis method, we test the consistency of a wide range of magnetic field models with RM maps of 11 extended sources in 5 clusters, for which the data were made available to us. We show that the data reveal no indication for a radial dependence of the average magnetic field strength, and in particular no indication for a correlation between the gas density and the field strength. The RM maps of a considerable fraction of the sources either require or are consistent with the presence of a spatially uniform magnetic field of a relatively small strength, $0.02\text{--}0.3\muG$, which contributes significantly to the RM. The RM maps of all but one source do not require a power-law magnetic field power spectrum, and most are consistent with a power spectrum dominated by a single wave length. The uncertainties in the magnetic field strengths (and spatial correlation lengths) derived from RM maps exceed an order of magnitude (and often more). These uncertainties imply, in particular, that there is no indication in current RM data for a systematic difference between the magnetic field strengths in radio-halo clusters and in radio-quiet clusters. With the improvement expected in the near future of the quality and quantity of RM data, our analysis method will enable one to derive more accurate constraints on magnetic fields in galaxy clusters.
\end{abstract}





\section{Introduction} \label{Sec:Intro}

The strength and spatial structure of magnetic fields in the intra-cluster medium (ICM) of galaxy clusters have been estimated using a wide range of methods. These methods include analyses of the properties of radio relic and halo sources, of cluster non-thermal X-ray emission, of cluster cold fronts (appearing in X-ray images), and of Faraday rotation measure maps of polarized radio sources \citep[for reviews, see][]{Carilli2002,Govoni2004}. The results of these analyses imply that the ICM of all galaxy clusters are permeated by  $\sim1\muG$ magnetic fields. Such fields play a critical role in determining the energy balance in the ICM plasma through their effects on heat conduction and cosmic ray propagation, with implications to many cluster phenomena \citep[for reviews, see][]{Feretti2012,Ferrari2008}.

In this paper we discuss the constraints that are implied by Faraday rotation measure (RM) data on the properties of the ICM magnetic fields. The Faraday effect causes a rotation of the plane of polarization of an electromagnetic wave, which is linearly proportional to the component of the magnetic field in the direction of propagation, chosen henceforth as $z$. The RM is defined as
\begin{equation} \label{Eq:RM}
\RM \equiv \dfracn{\lambda^2}{}{\psi} = \RM_0 \int \prnd{ \frac{B_{z}}{1\muG} }
			\prnd{ \frac{n_e}{10^{-3} \cm^{-3}} }
			\prnd{ \frac{\drm z}{1 \kpc} },
\end{equation}
where $\psi$ is the angle of rotation, $\lambda$ is the (radio) wavelength, $\RM_0\cong0.812\rad\m^{-2}$, $n_e$ is the (free) electron number density, and the integral is along the line of sight to the source \citep{Burn1966}. Cluster RM maps are obtained by fitting the wavelength dependence of the polarization angles measured along each line of sight to the $\lambda^2$ law of Eq.~(\ref{Eq:RM}).

The two dimensional RM maps do not allow one to uniquely determine the three dimensional magnetic field structure. Moreover, RM maps are available over only a small fraction of the angular extent of the clusters. Thus, RM constraints on the ICM magnetic fields are typically derived under simplifying assumptions regrading the magnetic field structure. Commonly used assumptions are that the magnetic field strength is correlated with the plasma density, and that the power spectrum of the spatial distribution of the magnetic field is a power-law in wave numbers. Furthermore, the range of model parameters (e.g. the range of wave numbers or the range of power-law indices) explored is limited in most derivations of magnetic field constraints, either by assumption or by some qualitative analysis of the two dimensional RM maps, in order to reduce the required computational resources.

The analysis presented here of the constraints imposed on the ICM magnetic fields improves on earlier work in several ways. First, we relax the commonly used assumptions of a correlation between the magnetic field strength and the plasma density, and of a power-law (in wave number) power spectrum. Indeed, we find that the RM maps show no evidence for a correlation between the field strength and the plasma density (as previously shown, e.g., by \citealt{Eilek2002}), and that a power-law power spectrum is generally not require to account for the observations. Second, we use a more efficient numerical analysis method, that allows us to explore a wide range of model parameters. Finally, we apply a uniform analysis to sources observed in several clusters, which allows us to draw conclusions applicable to the population of clusters (rather than to a single cluster or source).

RM maps have been derived and analyzed for radio sources in $16$ nearby galaxy clusters based on observations of more than $40$ radio sources (see Table~\ref{Tbl:GC}). The RM data available and earlier analyses of this data are described in \S~\ref{Sec:Prev}. Our analysis method is described in \S~\ref{Sec:Methods}, and our results are described in \S~\ref{Sec:Results}. We have applied our analysis method to all the data that were made available to us, 11 sources within 5 clusters, by F. Govoni, A. Bonafede, V. Vacca, and their collaborators. We thank them for making the data available, and note that our analysis method may be straightforwardly applied to all existing data. Our results are summarized and their implications are discussed in \S~\ref{Sec:Discussion}. In particular, we address the issue of magnetic field strength bi-modality between merging and non-merging galaxy clusters.

In our analysis we assume that the observed RM are entirely due to the ICM magnetic fields and not intrinsic to the sources \citep[e.g.][]{Pizzo2011}, and that the contribution of the magnetic field of our own Galaxy to the RM is smaller than the fit errors \citep[e.g.][]{Taylor1993}. We adopt a $\Lambda$CDM cosmology with $H_0=72\km\s^{-1}\Mpc^{-1}$, $\Omega_{\text{m}}=0.27$, and $\Omega_{\Lambda}=0.73$.



\section{Existing data \& earlier analyses} \label{Sec:Prev}


\subsection{Existing RM maps} \label{Sec:Obs}

\begin{figure*}[!htbp]
\centering
\includegraphics[width=0.7\linewidth,angle=0]{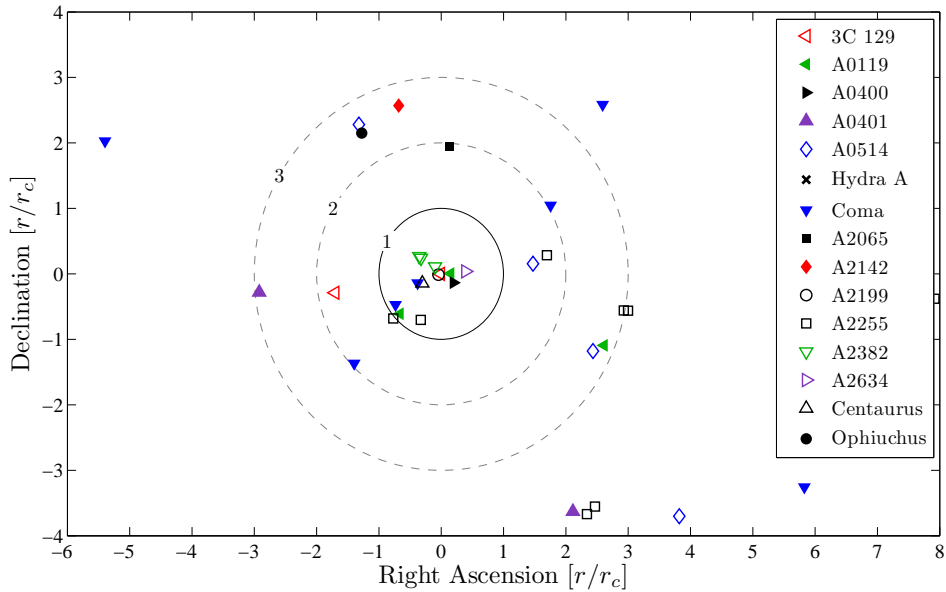}
\caption{Distribution of RM sources. Each marker represents the mean position of an extended RM source, relative to its host galaxy cluster (in the sense that the radio emission passes through its ICM) X-ray center. Note that a few of the central sources cover a relatively large area of their respective cores.\label{Fig:SrcDist}}
\end{figure*}

RM maps have been derived and analyzed for radio sources in $16$ nearby galaxy clusters (Table~\ref{Tbl:GC}). Detailed RM maps were constructed for more than $40$ radio sources (Table~\ref{Tbl:Src}) at different impact parameters (angular distance between the line of sight and the cluster X-ray center), lying within or behind the ICM. As can be seen in Figure~\ref{Fig:SrcDist}, the radio sources observed in these studies are distributed over a wide range of distances from their clusters' centers, with roughly half of the sources within $2r_c$ ($r_{c}$ as defined by Eq. \eqref{Eq:ne}).

\begin{table*}[!htbp]
{\centering
\caption{Galaxy Clusters used in RM studies\label{Tbl:GC}}
\begin{tabular}{ccccccccc}
\toprule
	\ColHead{Tbl:GC:Name}	{Cluster}	{}					&
	\ColHead{Tbl:GC:z}		{$z$}		{}					&
	\ColHead{Tbl:GC:Mrg}	{Mrg}		{}					&
	\ColHead{Tbl:GC:RH}		{Radio}		{}					&
	\ColHead{Tbl:GC:CC}		{CC}		{}					&
	\ColHead{Tbl:GC:Mass}	{$M$}		{10^{14}\Msun}		&
	\ColHead{Tbl:GC:beta}	{$\beta$}	{}					&
	\ColHead{Tbl:GC:rc}		{$r_c$}		{\!\kpc}			&
	\ColHead{Tbl:GC:ne}		{$n_0$}		{10^{-3}\cm^{-3}}	\\
\midrule
	3C 129		& $0.0210$	& \nodata		& \nodata	& \nodata	& $9.30^{+9.51}_{-3.85}$	& $0.601^{+0.260}_{-0.131}$	 & $224^{+125}_{-75}$
		& $2.1^{+0.8}_{-0.5}$	\\
	A0119		& $0.0442$	& \checkmark	& \nodata	& \nodata	& $10.76^{+1.50}_{-1.39}$	& $0.675^{+0.026}_{-0.023}$	 & $356^{+19}_{-18}$
		& $1.8^{+0.1}_{-0.1}$	\\
	A0400		& $0.0244$	& \checkmark	& \nodata	& \nodata	& $2.07^{+0.30}_{-0.25}$	& $0.534^{+0.014}_{-0.013}$	 & $108^{+6}_{-6}$
		& $2.4^{+0.1}_{-0.1}$	\\
	A0401		& $0.0737$	& \checkmark	& G			& \nodata	& $16.59^{+1.62}_{-1.62}$	& $0.613^{+0.010}_{-0.010}$	 & $177^{+8}_{-7}$
		& $7.1^{+0.6}_{-0.5}$	\\
	A0514\tnt{Tbl:GC:Note:A0514}
				& $0.0713$	& \checkmark	& \nodata	& \nodata	& $8.94^{+1.82}_{-1.45}$	& $0.600^{+0.120}_{-0.070}$	 & $411^{+144}_{-94}$
					& $0.5^{+0.1}_{-0.1}$	\\
	Hydra A		& $0.0539$	& \nodata		& \nodata	& W			& $5.94^{+0.91}_{-0.84}$	& $0.573^{+0.003}_{-0.003}$	 & $36^{+1}_{-1}$
		& $42.4^{+4.4}_{-4.0}$	\\
	Coma		& $0.0231$	& \checkmark	& G+R		& \nodata	& $19.38^{+2.08}_{-1.97}$	& $0.654^{+0.019}_{-0.021}$	 & $241^{+15}_{-14}$
		& $3.6^{+0.7}_{-0.7}$	\\
	A2065		& $0.0726$	& \checkmark	& \nodata	& W			& $23.37^{+29.87}_{-9.42}$	& $1.162^{+0.734}_{-0.282}$	 & $497^{+260}_{-134}$
		& $2.4^{+0.8}_{-0.5}$	\\
	A2142		& $0.0909$	& \nodata		& M			& W			& $21.04^{+5.46}_{-3.69}$	& $0.591^{+0.006}_{-0.006}$	 & $111^{+4}_{-4}$
		& $18.9^{+0.8}_{-0.8}$	\\	
	A2199		& $0.0302$	& \nodata		& \nodata	& S			& $6.73^{+0.52}_{-0.51}$	& $0.655^{+0.019}_{-0.021}$	 & $98^{+7}_{-6}$
		& $9.9^{+0.4}_{-0.4}$	\\
	A2255		& $0.0654$	& \checkmark	& G+R		& \nodata	& $18.65^{+2.01}_{-1.67}$	& $0.797^{+0.033}_{-0.030}$	 & $429^{+25}_{-23}$
		& $2.1^{+0.2}_{-0.2}$	\\
	A2382\tnt{Tbl:GC:Note:A2382}
				& $0.0618$	& \nodata		& \nodata	& \nodata	& $6.7^{+1.1}_{-1.1}$		& $0.900^{+0.700}_{-0.200}$	 & $370^{+250}_{-128}$
		& $1.3^{+0.3}_{-0.2}$	\\	
	A2634		& $0.0314$	& \nodata		& \nodata	& W			& $5.35^{+1.34}_{-1.04}$	& $0.640^{+0.051}_{-0.043}$	 & $257^{+30}_{-27}$
		& $2.9^{+0.2}_{-0.2}$	\\
	Centaurus	& $0.0114$	& \nodata		& \nodata	& S			& $3.78^{+0.23}_{-0.18}$	& $0.495^{+0.011}_{-0.010}$	 & $29^{+3}_{-3}$
		& $20.9^{+1.5}_{-1.4}$	\\
	Ophiuchus	& $0.0280$	& \checkmark	& M			& \nodata	& $32.43^{+4.04}_{-3.38}$	& $0.747^{+0.035}_{-0.032}$	 & $196^{+16}_{-15}$
		& $8.1^{+0.5}_{-0.5}$	\\
\bottomrule
\end{tabular}
\\}
{\footnotesize
\begin{enumerate}
	\labitem{\dag}{Tbl:GC:Note:A0514}	Mass ref.: \citealt{Girardi1998}, $\beta$-model ref.: \citealt{Govoni2001}.
	\labitem{\ddag}{Tbl:GC:Note:A2382}	Mass ref.: \citealt{Demarco2003}, $\beta$-model ref.: \citealt{Guidetti2008}.
\end{enumerate}
\textsc{Tab}. \ref{Tbl:GC}.---
	Galaxy Clusters used in RM studies, including some relevant cluster properties.
	\begin{inparaenum}[(1)]
		\item \label{Tbl:GC:Name}	Cluster name;
		\item \label{Tbl:GC:z}		Redshift (\textsc{Nasa/ipac  Extragalactic  Database});
		\item \label{Tbl:GC:Mrg}	Merger activity \citep[and references therein]{Bonafede2011};
		\item \label{Tbl:GC:RH}		Diffuse radio emission: G = giant halo, M = mini halo, R = relic \citep{Feretti2012};
		\item \label{Tbl:GC:CC}		Cool-Core: W = weak, S = strong \citep{Hudson2010};
		\item \label{Tbl:GC:Mass}	Virial mass \citep[unless otherwise noted]{Reiprich2002};
		\item \label{Tbl:GC:beta}
		\item \label{Tbl:GC:rc}
		\item \label{Tbl:GC:ne}		$\beta$-model (Eq.~\ref{Eq:ne}) parameters with 1-$\sigma$ fit errors adapted according to our adopted cosmology \citep[unless otherwise noted]{Chen2007}.
	\end{inparaenum}
}\\
\end{table*}

\begin{table*}[!htbp]
{\centering
\caption{Radio sources analyzed in previous work\label{Tbl:Src}}
\begin{tabular}{cccccccc}
\toprule
	\ColHead{Tbl:Src:GC}	{Cluster}			{}	&
	\ColHead{Tbl:Src:Name}	{Source}			{}	&
	\ColHead{Tbl:Src:Morph}	{Morph.}			{}	&
	\ColHead{Tbl:Src:FR}	{FR}				{}	&
	\ColHead{Tbl:Src:Class}	{Class}				{}	&
	\ColHead{Tbl:Src:b}		{$r_{\bot}/r_c$}	{}	&
	\ColHead{Tbl:Src:LOS}	{LOS}				{}	&
	\ColHead{Tbl:Src:Ref}	{Ref.}				{}	\\
\midrule
		3C 129		& 3C 129			& HT		& \nodata	& galaxy	& 1.73	& $0$		& \ref{Tbl:Src:Ref:Taylor2001}		\\
					& 3C 129.1			& \nodata	& I			& galaxy	& 0.01	& $0$		& \ref{Tbl:Src:Ref:Taylor2001}		\\
		A0119		& 0053-015			& NAT		& \nodata	& E			& 0.15	& $0$		& \ref{Tbl:Src:Ref:Feretti1999}		\\
					& 0053-016			& NAT		& \nodata	& galaxy	& 0.90	& $0$		& \ref{Tbl:Src:Ref:Feretti1999}		\\
					& 3C 029			& \nodata	& I			& S			& 2.83	& $0$		& \ref{Tbl:Src:Ref:Feretti1999}		\\
		A0400		& 3C 075			& T			& \nodata	& E0		& 0.24	& $0$		& \ref{Tbl:Src:Ref:Eilek2002}		\\
		A0401		& A401A				& NAT		& \nodata	& galaxy	& 2.93	& $0$		& \ref{Tbl:Src:Ref:Govoni2010}		\\
					& A401B				& HT		& \nodata	& galaxy	& 4.20	& $0$		& \ref{Tbl:Src:Ref:Govoni2010}		\\
		A0514		& A514A				& \nodata	& II		& \nodata	& 2.63	& $\infty$	& \ref{Tbl:Src:Ref:Govoni2001}		\\
					& A514B2			& NAT		& \nodata	& D			& 1.48	& $0$		& \ref{Tbl:Src:Ref:Govoni2001}		\\
					& A514C				& \nodata	& II		& \nodata	& 4.53	& $\infty$	& \ref{Tbl:Src:Ref:Govoni2001}		\\
					& A514D				& T			& I			& galaxy	& 2.71	& $0$		& \ref{Tbl:Src:Ref:Govoni2001}		\\
					& A514E				& \nodata	& \nodata	& QSO		& 5.32	& $\infty$	& \ref{Tbl:Src:Ref:Govoni2001}		\\
		Hydra A		& Hydra A			& \nodata	& \nodata	& galaxy	& 20.0	& $0$		& \ref{Tbl:Src:Ref:Taylor1993}		\\
		\textit{Coma}
					& 5C 04.042			& \nodata	& II		& galaxy	& 5.77	& $\infty$	& \ref{Tbl:Src:Ref:Bonafede2010}	\\
					& 5C 04.074\tnt{Tbl:Src:Note:edge}
										& \nodata	& II		& \nodata	& 1.95	& $\infty$	& \ref{Tbl:Src:Ref:Bonafede2010}	\\
					& 5C 04.081\tnt{Tbl:Src:Note:halo}
										& NAT		& I			& cD		& 0.87	& $0$		& \ref{Tbl:Src:Ref:Bonafede2010}	\\
					& 5C 04.085\tnt{Tbl:Src:Note:halo}
										& WAT		& I-II		& cD		& 0.40	& $0$		& \ref{Tbl:Src:Ref:Bonafede2010}	\\
					& 5C 04.114\tnt{Tbl:Src:Note:edge}
										& \nodata	& I			& \nodata	& 2.04	& $\infty$	& \ref{Tbl:Src:Ref:Bonafede2010}	\\
					& 5C 04.127			& \nodata	& \nodata	& QSO		& 3.66	& $\infty$	& \ref{Tbl:Src:Ref:Bonafede2010}	\\
					& 5C 04.152			& \nodata	& II		& galaxy	& 6.68	& $\infty$	& \ref{Tbl:Src:Ref:Bonafede2010}	\\
		\textit{A2065}	
					& A2065A			& \nodata	& II		& galaxy	& 1.96	& $\infty$	& \ref{Tbl:Src:Ref:Govoni2010}		\\
		\textit{A2142}
					& A2142A			& HT		& I			& galaxy	& 2.66	& $0$		& \ref{Tbl:Src:Ref:Govoni2010}		\\
		\textit{A2199}
					& 3C 338			& \nodata	& I			& cD2		& 0.05	& $\infty$	& \ref{Tbl:Src:Ref:Vacca2012}		\\
		A2255		& Double\tnt{Tbl:Src:Note:edge}
										& HT		& \nodata	& galaxy	& 2.98	& \nodata	& \ref{Tbl:Src:Ref:Pizzo2011}		\\
					& Goldfish\tnt{Tbl:Src:Note:halo}
										& NAT		& \nodata	& galaxy	& 1.72	& \nodata	& \ref{Tbl:Src:Ref:Pizzo2011}		\\
					& TRG\tnt{Tbl:Src:Note:halo}
										& NAT		& \nodata	& galaxy	& 1.03	& \nodata	& \ref{Tbl:Src:Ref:Pizzo2011}		\\
					& Bean				& HT		& \nodata	& galaxy	& 11.6	& \nodata	& \ref{Tbl:Src:Ref:Pizzo2011}		\\
					& Embryo			& WAT		& \nodata	& 			& 7.93	& \nodata	& \ref{Tbl:Src:Ref:Pizzo2011}		\\
					& Beaver			& NAT		& \nodata	& galaxy	& 4.33	& \nodata	& \ref{Tbl:Src:Ref:Pizzo2011}		\\
					& F1, F2, F3\tnt{Tbl:Src:Note:halo}
										& \nodata	& \nodata	& filaments	& 		& \nodata	& \ref{Tbl:Src:Ref:Pizzo2011}		\\
					& J1712.4+6401\tnt{Tbl:Src:Note:halo}
										& \nodata	& \nodata	& \nodata	& 0.78	& \nodata	& \ref{Tbl:Src:Ref:Govoni2006}		\\
					& J1713.3+6347		& \nodata	& \nodata	& \nodata	& 4.35	& \nodata	& \ref{Tbl:Src:Ref:Govoni2006}		\\
					& J1713.5+6402\tnt{Tbl:Src:Note:edge}
										& \nodata	& \nodata	& \nodata	& 3.05	& \nodata	& \ref{Tbl:Src:Ref:Govoni2006}		\\
					& J1715.1+6402		& \nodata	& \nodata	& \nodata	& 8.14	& \nodata	& \ref{Tbl:Src:Ref:Govoni2006}		\\
		A2382		& PKS 2149-158 A	& \nodata	& I			& E			& 0.44	& $0$		& \ref{Tbl:Src:Ref:Guidetti2008}	\\
					& PKS 2149-158 B	& \nodata	& I			& E			& 0.41	& $0$		& \ref{Tbl:Src:Ref:Guidetti2008}	\\
					& PKS 2149-158 C	& \nodata	& I			& E			& 0.15	& $0$		& \ref{Tbl:Src:Ref:Guidetti2008}	\\
		A2634		& 3C 465			& T			& I			& cD		& 0.39	& $0$		& \ref{Tbl:Src:Ref:Eilek2002}		\\
		Centaurus	& PKS 1246-410		& \nodata	& I			& cD1		& 0.34	& $0$		& \ref{Tbl:Src:Ref:Taylor2002}		\\
		\textit{Ophiuchus}
					& OPHIB				& HT		& I			& galaxy	& 2.50	& $0$		& \ref{Tbl:Src:Ref:Govoni2010}		\\
\bottomrule
\end{tabular}
\\}
{\footnotesize
\begin{enumerate}
	\labitem{\dag}{Tbl:Src:Note:halo}	The radio source is coincident with the cluster RH.
	\labitem{\ddag}{Tbl:Src:Note:edge}	The radio source might be coincident with the cluster RH (edge of diffuse emission).
\end{enumerate}
\textsc{Tab}. \ref{Tbl:Src}.---
	The radio sources for which RM maps have been derived and analyzed in previous works, along with some of their relevant characteristics.
	In this work we analyzed the sources in Coma, A2065, A2142, A2199, and Ophiuchus (italicized).
	\begin{inparaenum}[(1)]
		\item \label{Tbl:Src:GC}		Host GC name;
		\item \label{Tbl:Src:Name}	Source name;
		\item \label{Tbl:Src:Morph}	Radio morphology: (H)T = (Head) Tail, (N/W)AT = (Narrow/Wide) Angle Tail;
		\item \label{Tbl:Src:FR}		Fanaroff-Riley classification;
		\item \label{Tbl:Src:Class}	Source object type;
		\item \label{Tbl:Src:b}		Impact parameter relative to X-ray center;
		\item \label{Tbl:Src:LOS}		Line of sight position: $\infty$ = background source, $0$ = cluster member;
		\item \label{Tbl:Src:Ref}		RM reference:
		\begin{inparaenum}[1.]
			\item \label{Tbl:Src:Ref:Bonafede2010}	\citealt{Bonafede2010};
			\item \label{Tbl:Src:Ref:Eilek2002}		\citealt{Eilek2002};
			\item \label{Tbl:Src:Ref:Feretti1999}	\citealt{Feretti1999};
			\item \label{Tbl:Src:Ref:Govoni2001}	\citealt{Govoni2001};
			\item \label{Tbl:Src:Ref:Govoni2006}	\citealt{Govoni2006};
			\item \label{Tbl:Src:Ref:Govoni2010}	\citealt{Govoni2010};
			\item \label{Tbl:Src:Ref:Guidetti2008}	\citealt{Guidetti2008};
			\item \label{Tbl:Src:Ref:Perley1991}	\citealt{Perley1991};
			\item \label{Tbl:Src:Ref:Pizzo2011}		\citealt{Pizzo2011};
			\item \label{Tbl:Src:Ref:Taylor1993}	\citealt{Taylor1993};
			\item \label{Tbl:Src:Ref:Taylor2001}	\citealt{Taylor2001};
			\item \label{Tbl:Src:Ref:Taylor2002}	\citealt{Taylor2002};
			\item \label{Tbl:Src:Ref:Vacca2010}		\citealt{Vacca2010};
			\item \label{Tbl:Src:Ref:Vacca2012}		\citealt{Vacca2012}.
		\end{inparaenum}
	\end{inparaenum}
}\\
\end{table*}

The sources included in our study are indicated in Table~\ref{Tbl:Src}. More than half (7/11) of our analyzed sources are within Coma, a prominent radio halo (RH) cluster, and 4 of these sources are coincident with the RH. We also have one source in each of the clusters A2065, A2142, A2199 and Ophiuchus, none of which are hosting a RH.

\subsection{Earlier analyses} \label{Sec:PrevWork}

For five of the galaxy clusters listed in Table~\ref{Tbl:GC} (Coma, A2199, Hydra~A, A2255, and A2382), detailed models of the ICM magnetic fields have been constructed and constrained by the observed RM maps \citep{Taylor1993,Govoni2006,Guidetti2008,Bonafede2010,Vacca2012}. For the rest of the sources only a limited analysis was carried out \citep{Bonafede2011,Govoni2010}. We briefly summarize below the main results of both types of studies.


\subsubsection{Clusters for which detailed analyses were carried out}

For Coma, \cite{Bonafede2010} assumed that the spatial distribution of the magnetic field is given by a realization of a Gaussian random field with a Fourier power spectrum that is a power-law of the wave numbers. They have assumed a Kolmogorov spectral index, $n=11/3$, and by comparing various statistical properties of the observed RM images and those simulated, the spatial wavelength range was constrained to $\Lambda_{\text{min}}=2\kpc$ and $\Lambda_{\text{max}}=34\kpc$.

In order to generate a correlation of the field strength with the plasma density, the (real space) magnetic field obtained in the realizations was multiplied by an $r$-dependent factor to generate
\begin{equation}
\mean{B\of{r}} = \Bc \psqr{n_e\of{r}/n_0}^{\eta},
\end{equation}
where the average is over all realizations.
The RM maps derived from the magnetic field realizations were compared to the observed RM maps, to constrain the central magnetic field strength $\Bc$ to be in the range $3.9\text{--}5.4\muG$, and the correlation index $\eta$ to be in the range $0.4\text{--}0.7$ ($1\sigma$ c.l.).

A similar analysis has been carried out for A2199 by \cite{Vacca2012}. In their analysis, $\Lmax=35\kpc$, $\Lmin=0.7\kpc$, and $n=2.8$ were used, based on fitting a power-law spectrum to the 2D RM map (from which $\Lmax=35\pm28\kpc$, $\Lmin=0.7\pm0.1\kpc$, and $n=2.8\pm1.3$ were inferred). Using these values and following a method similar to that of \cite{Bonafede2010} for Coma, they find $\Bc=11.7\pm9.0\muG$ and $\eta=0.9\pm0.5$ ($1\sigma$ c.l.).

Besides Coma and Abell~2199, the ICM magnetic fields of only three other clusters -- Hydra~A \citep{Taylor1993}, A2255 \citep{Govoni2006}, and A2382 \citep{Guidetti2008} -- have been investigated with a similar approach, i.e. by simulating random magnetic field models and constraining their properties by comparison to the RM data.

For Hydra~A, several models were explored for the ICM fields. \cite{Taylor1993} find $\sim20\muG$ to $\sim40\muG$ magnetic field strength assuming spatial variability on $\sim4\kpc$ scale, and some evidence for magnetic fields with cluster scale ordering.

For Abell~2255, a RH cluster, \cite{Govoni2006} estimate the cluster core field strength to be $\sim2.5\muG$. They try several models with a few different values for the power-spectrum power-law index, keeping the field strength at the cluster center constant at $2-2.5\muG$, and conclude that it is necessary to use a (spatially) variable power index (in the range $n=2-4$).

For Abell~2382, \cite{Guidetti2008} produce a fit with central magnetic field strength, $\Bc=\range{1}{13}\muG$ ($1\sigma$ confidence region) for models with different correlation index between the average magnetic field strength and the gas density.
Note that all their models produce similar magnetic field strengths when averaged over a large volume ($1\Mpc^3$).


\subsubsection{Clusters with limited analyses} \label{Sec:GaussCell}

Comparing the standard deviation of each RM source, $\sRM$ (averaging the observed values over all valid pixels in each source), to the distance from the cluster center, not normalized to account for the different cluster sizes (e.g. to $r_c$, the core radius of the $\beta$-model), and assuming a "Gaussian cell model\footnote{an analytical formulation based on the approximation that the magnetic field strength is constant throughout the cluster and that its direction varies over a single length scale}", \cite{Govoni2010} find these galaxy clusters are consistent with $\Bc\sim1\text{--}4\muG$.


\section{Method} \label{Sec:Methods}

We assume that the cluster's magnetic field may be described as a realization of a random process, the statistical properties of which we seek to constrain using the measured RM map. We consider a statistical model acceptable if the power spectrum of the observed (two dimensional) RM map is consistent with being drawn from the distribution of power spectra predicted by the model. To determine whether or not this is the case, we compare the distribution of some measure of the differences between the RM power spectra of different realizations of the model with the distribution of the same measure of differences between the model's power spectra and the observed power spectrum. We define the RM power spectrum as
\begin{equation}
\RM\of{k}^2=\mean{\abs{\RM\of{k_x,k_y}}^2}_{\abs{\prnd{k_x,k_y}}=k},
\end{equation}
where $\prnd{k_x,k_y}$ is the 2D Fourier $k$-vector corresponding to the projected distance on the plane of the sky, and the average on the rhs is over all vectors with magnitude $k$. For our analysis we use two measures of the distance between two power spectra, $\RM_1(k)$ and $\RM_2(k)$: the sum and the maximum of the absolute value of the logarithm of the ratio of powers in the two power spectra at all wave numbers, i.e. $\sum_k\abs{\log(\RM_1(k)/\RM_2(k))}$ and $\max_k\abs{\log(\RM_1(k)/\RM_2(k))}$, respectively. For each statistical model we generated $\sim500$ realizations and compared the resulting distance measure distributions using the Kolmogorov-Smirnov (KS) test. Similar results are obtained for the two distance measures used.

We write the magnetic field as a sum of Fourier components, $\sum_\mathbf{k}\tilde{\mathbf{B}}(\mathbf{k})\exp(i\mathbf{k}\cdot\mathbf{x})$ with
\begin{equation} \label{Eq:Bk}
\tilde{\mathbf{B}}\of{\kvec}
		= B_\mathbf{k} e^{i \bOm}
				\psqr{ \khat_1 \cos\of{\aOm}+ \khat_2 \sin\of{\aOm}},
\end{equation}
where $\khat_1$ and $\khat_2$ are two unit vectors
orthogonal to $\khat$ (chosen to align along the $z$-direction of Cartesian coordinates) and to each other (ensuring $\Nabla\cdot\Bvec=0$), and $B_\mathbf{k}$ is real. Since the magnetic field is assumed to be statistically homogeneous and isotropic, which choose the phases $\aOm$ and $\bOm$ to be uniformly distributed in $\psqr{-\pi,+\pi}$. The $k$-space resolution is chosen as $\sim2\pi/r_c$.

We consider two types of power spectra for the amplitude $B_\mathbf{k}$:
a single characteristic wave length $\Lambda$, with $B_\mathbf{k}=0$ for $k\neq 2\pi/\Lambda$, and a power-law spectrum with index $n$, with $k^2 B_\mathbf{k}^2\propto k^{-n}$ for $k$ in some fixed range $\psqr{\kmin,\kmax}$ (corresponding to $\Lmin=1\kpc,\Lmax=32\kpc$) and zero elsewhere. In both cases, we allow the presence of a uniform ($k=0$) component (as proposed by \citealt{Taylor1993}), with amplitude $\Bmsk$. The average magnetic field strength, $\Bc\equiv(3\mean{B_z^2})^{1/2}$, is assumed to be constant throughout the cluster (i.e., with no radial dependance). As described in \S~\ref{Sec:Results}, the simple models we consider are sufficient to account for the vast majority of observations, and hence more general models are not motivated by the data.

For any realization of the magnetic field model, the Faraday RM map is obtained using Eq.~(\ref{Eq:RM}) and a $\beta$-model density profile,
\begin{equation}\label{Eq:ne}
n_e = n_0 \prnd{1 + r^2 / r_c^2}^{-\frac{3}{2}\beta},
\end{equation}
from the literature (as listed in Table~\ref{Tbl:GC}). We treat cluster members as situated precisely halfway through the ICM, and other sources as lying at infinity. For these two cases, the integral of Eq.~(\ref{Eq:RM}) may be solved analytically, as shown in Appendix~\ref{Apx:B2RM}. Note also that due to the linear nature of Eq.~\ref{Eq:RM}, it is possible to explore as many spectral index values as desired, without having to simulate the magnetic field more than once at each wave number $k$.

In order to account for the finite angular resolution of the observations, RM values at each pixel are obtained by properly averaging the values obtained in a band of surrounding pixels corresponding to the reported angular beam size. Our results are not sensitive to $\sim10$\% changes in the beam size adopted for the calculation. RM values are derived in the simulations only at coordinates specified by pixels with observed RM data, and at a band of pixels surrounding them. It is thus not necessary to use a large cubic Fourier grid and not necessary to use heavily resource consuming FFT calculations.

Finally, after the convolution with the beam, Gaussian noise is added, with standard deviation equal to the mean observed RM fit errors. Our realizations showed that the noise has only a little influence on the RM power-spectrum. Some contribution to the power on short wavelengths is obtained only when the level of the noise is increased to $\sim100\%$.


\section{Results} \label{Sec:Results}

\begin{figure*}[p]
\centering
\includegraphics[width=0.9\linewidth, trim=0cm 0cm 0cm 0.45cm]{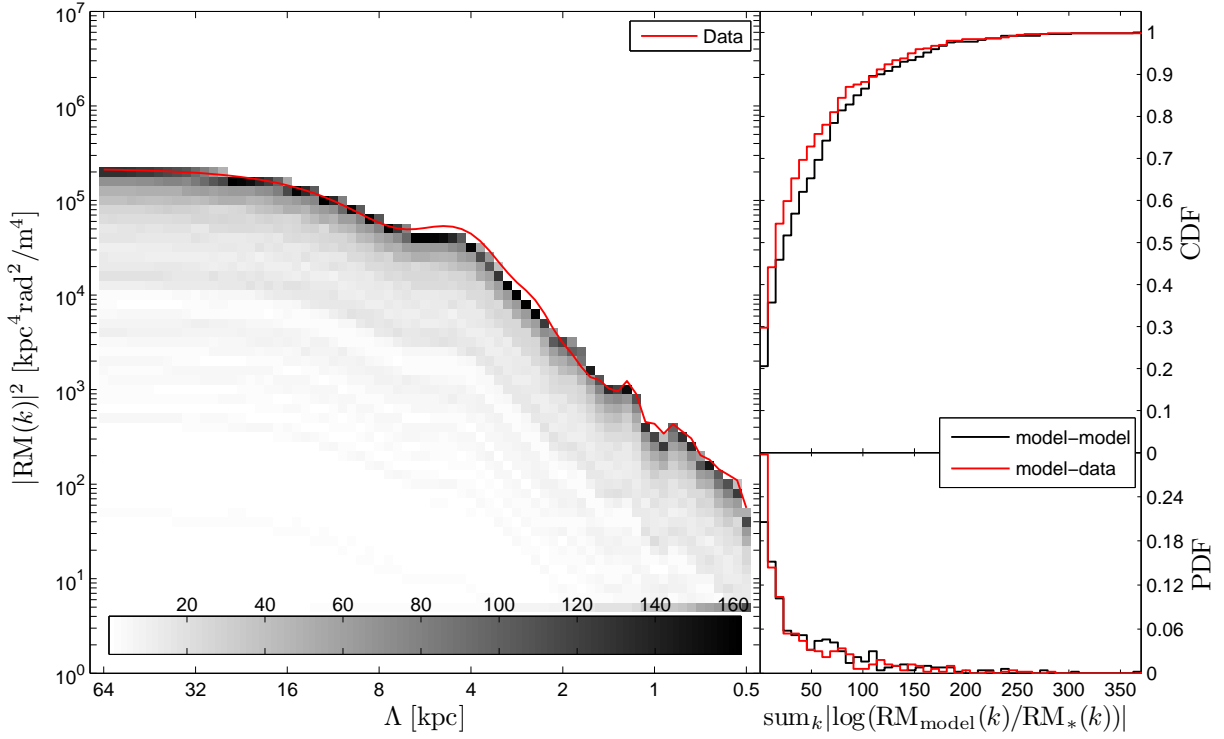}
\caption{A homogeneous field model for 5C4.114. Left panel: a histogram of realization power-spectra, for a homogeneous field model with $\Bmsk=26.1\unit{nG}$, compared to the data power-spectrum (red). Colors indicate the number of realizations for which the various values were obtained. Right panels: the (binned) CDF and PDF (top and bottom respectively) of the sum of absolute logarithmic differences between the data to the model (red) and between realizations (black). The maximized KS test statistic is $p_{\text{KS}}=0.016$. \label{Fig:MskCom114}}
\end{figure*}

\begin{figure*}[p]
\centering
\includegraphics[width=0.9\linewidth, trim=0cm 0cm 0cm 0.45cm]{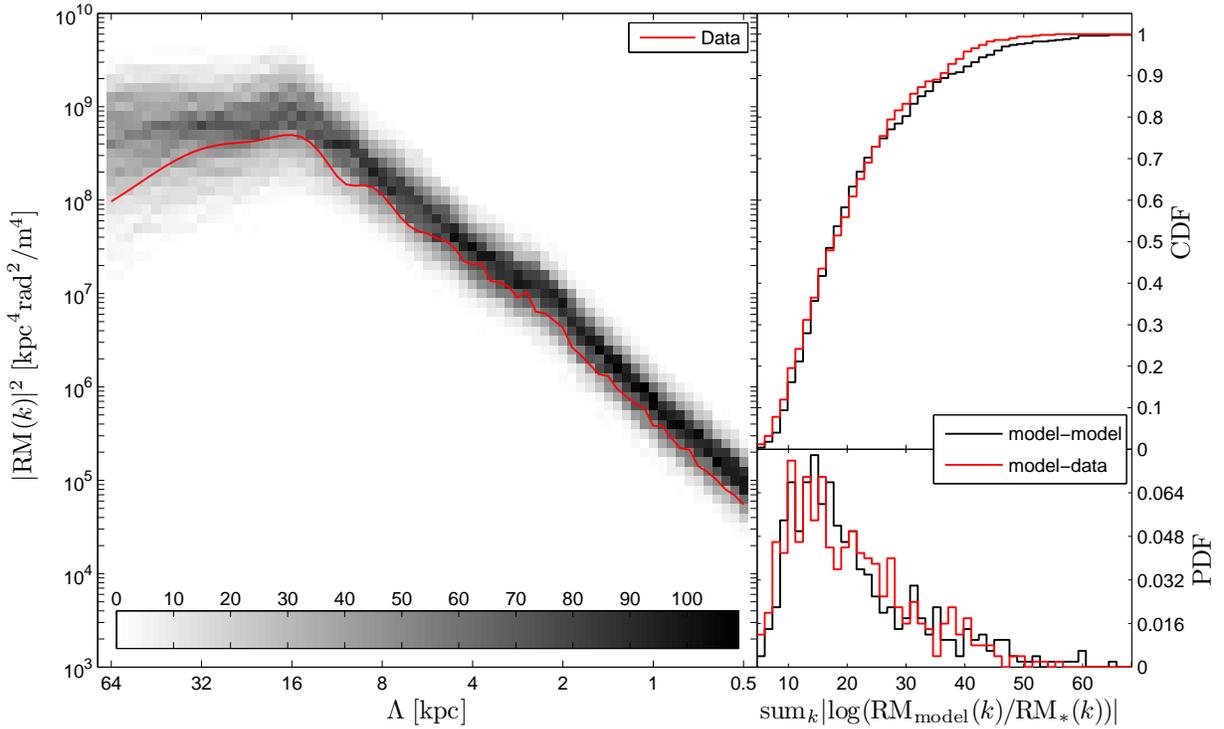}
\caption{Same as Figure~\ref{Fig:MskCom114}, but for the single-scale field model with $\Lambda=16\kpc$ and $B_{0}=(\mean{3B_z^2})^{1/2}=0.964\muG$ for A2199A (without a homogeneous component). The maximized KS test statistic is $p_{\text{KS}}=0.762$. \label{Fig:SngA2199A}}
\end{figure*}

In this section we present the results of our analysis, which was described in \S~\ref{Sec:Methods}. For every source we simulated five types of magnetic field models: single-scale, power-law, both with and without a homogeneous component, and a simple homogeneous field. Figures~\ref{Fig:MskCom114} and~\ref{Fig:SngA2199A} show the RM power spectra of 5C4.114 and A2199A, and compare them to those obtained for two types of models: a homogeneous field model for 5C4.114 and a single scale model for A2199A. The results are summarized in Table~\ref{Tbl:Results}, where $95\%$ c.l. ranges are shown for accepted model parameters (as is the convention also in what follows).

For some of the sources, most notably 5C4.114 and A2142A, the RM power-spectra resemble closely those obtained for a homogeneous magnetic field, reflecting the fact that the power spectrum is determined by the shape of the sources. Our realizations show that a homogeneous sub $\!\muG$ magnetic field is sufficient in order to generate their observed RM power-spectra: $\range{0.026}{0.030}\muG$ for 5C4.114 and $\range{0.18}{0.20}\muG$ for A2142A.

\begin{figure}[!hb]
\centering
\includegraphics[width=\linewidth,angle=0]{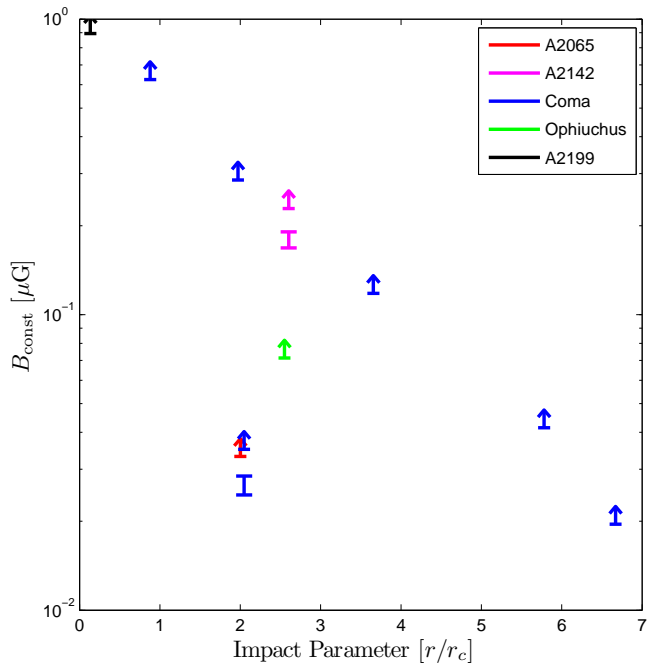}
\caption{The lower-bounds on the magnetic field, as derived using Eq.~\eqref{Eq:MinB0} (arrows), and the ranges of magnetic field strengths in the homogeneous field models accounting for the A2142A and 5C4.114 RM data (95\% c.l. error bars).  The small offset between the bounds and allowed ranges is due to the fact that the lower bounds are derived based on the RM of a single pixel, in contrast with the (weighted) averaging over many pixels inherent to our statistical analysis.\label{Fig:Msk}}
\end{figure}

Figure~\ref{Fig:Msk} shows the minimal magnitude of the magnetic field implied by observations of each source, which is the minimal magnitude of a homogeneous field required to generate the maximal (absolute) RM value of each source. Using Eqs. (\ref{Eq:RM}) and (\ref{Eq:ne}), this minimal field is
\begin{equation} \label{Eq:MinB0}
B_0 \ge c_{\text{LOS}} \sqrt{3} \max_{x,y} \frac {\abs{\RM\of{x,y}}\prnd{1+\frac{x^2+y^2}{r_c^2}}^{\frac{3}{2}\beta-\frac{1}{2}}\Gamma\of{\frac{3}{2}\beta}}
	{\RM_0 n_0 r_c \sqrt{\pi}\Gamma\of{\frac{3}{2}\beta-\frac{1}{2}}},
\end{equation}
where $c_{\text{LOS}}=1$ for a background source or $c_{\text{LOS}}=2$ for a cluster member, and $\prnd{x,y}$ correspond to right ascension and declination relative to the cluster center. The lower limits are are $\sim30\%$ higher than the acceptable ranges of values inferred above for 5C4.114 and A2142A. The small offset supports the validity of our simulations' results. It is due to the fact that the lower limits are derived based on the RM of a single pixel, in contrast with the (weighted) averaging over many pixels inherent to our statistical analysis.

\begin{figure*}[!t]
\centering
\begin{tabular}{cc}
\includegraphics[width=0.48\linewidth]{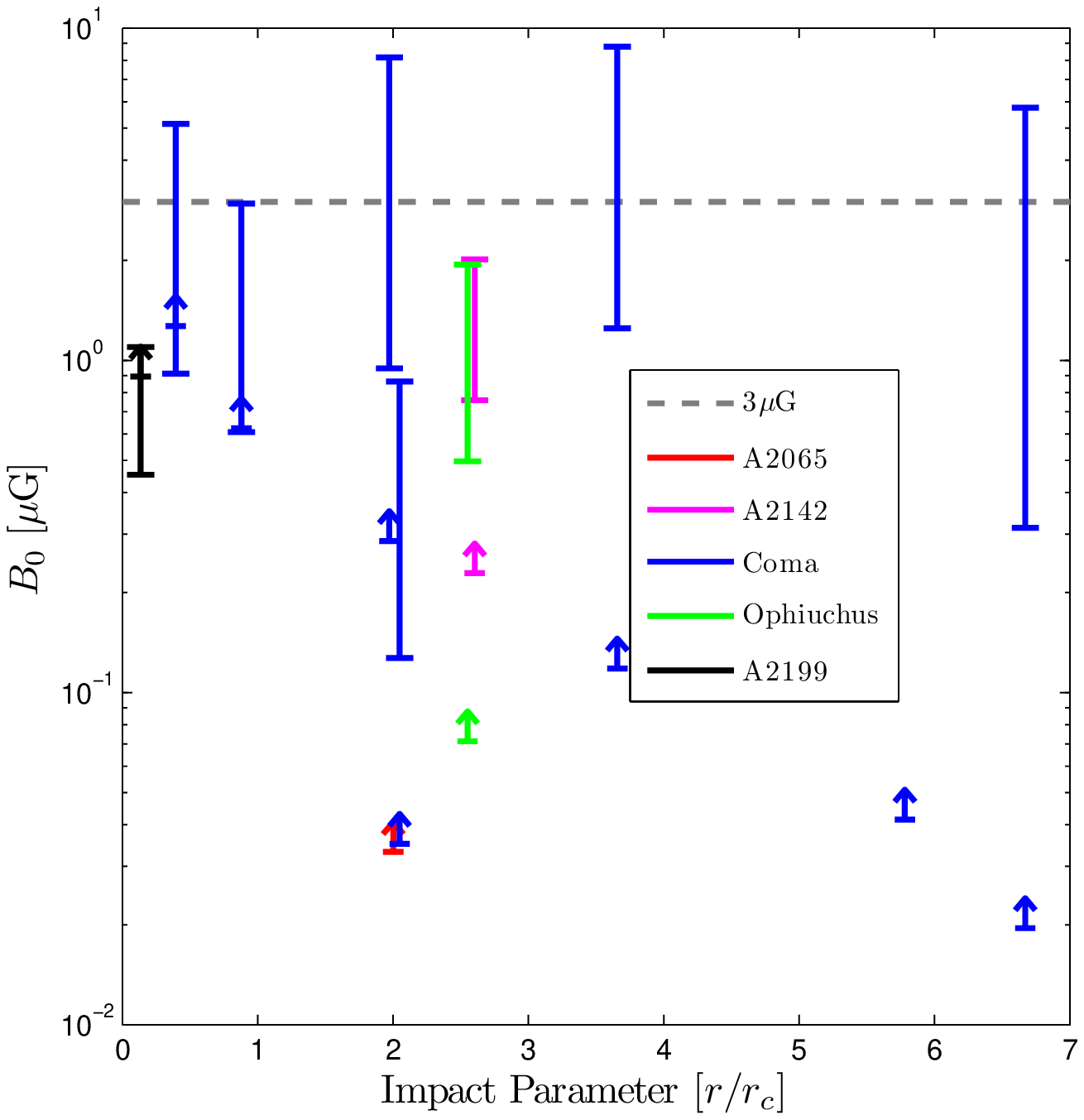} &
\includegraphics[width=0.48\linewidth]{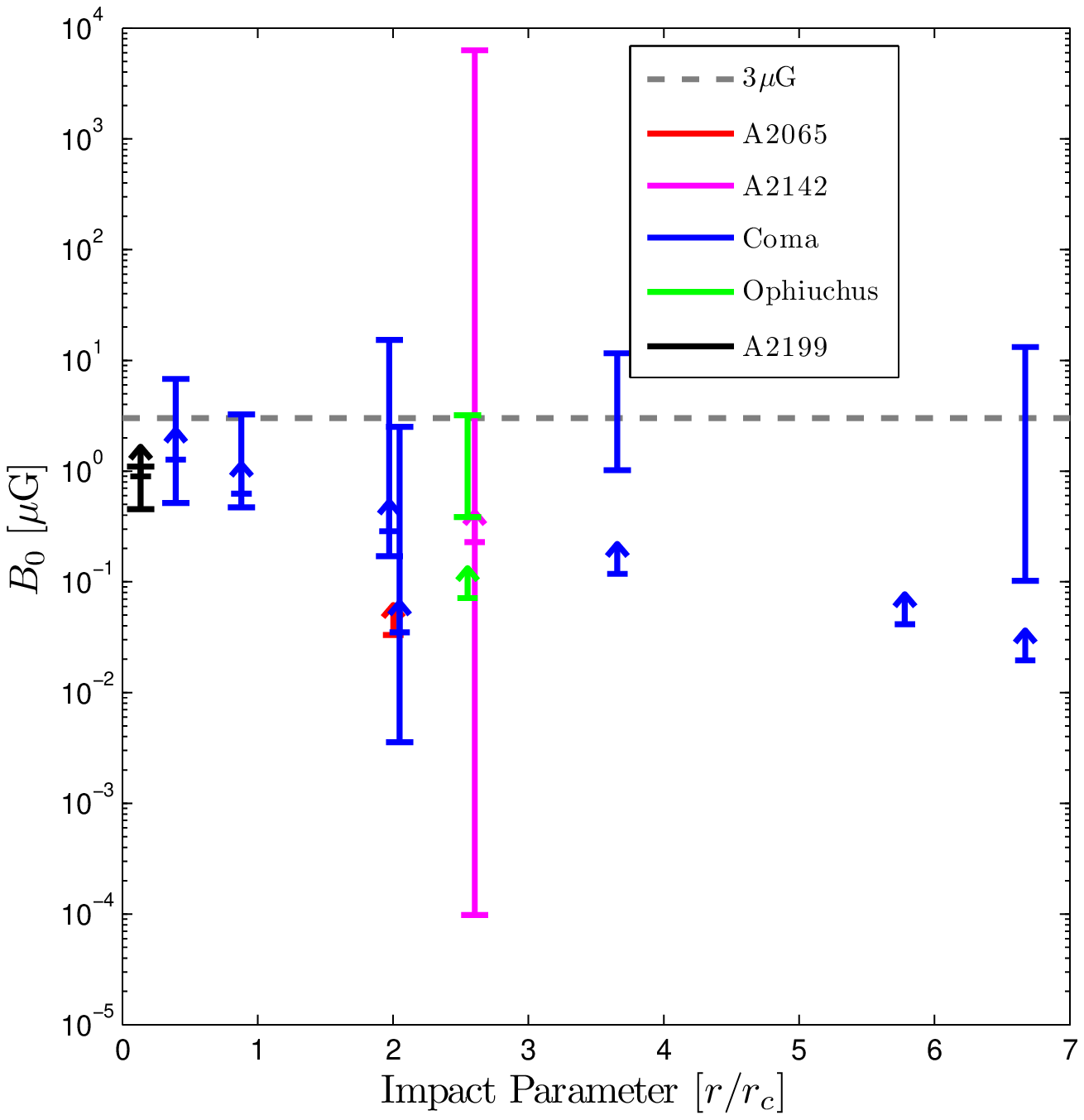} \\
\includegraphics[width=0.48\linewidth]{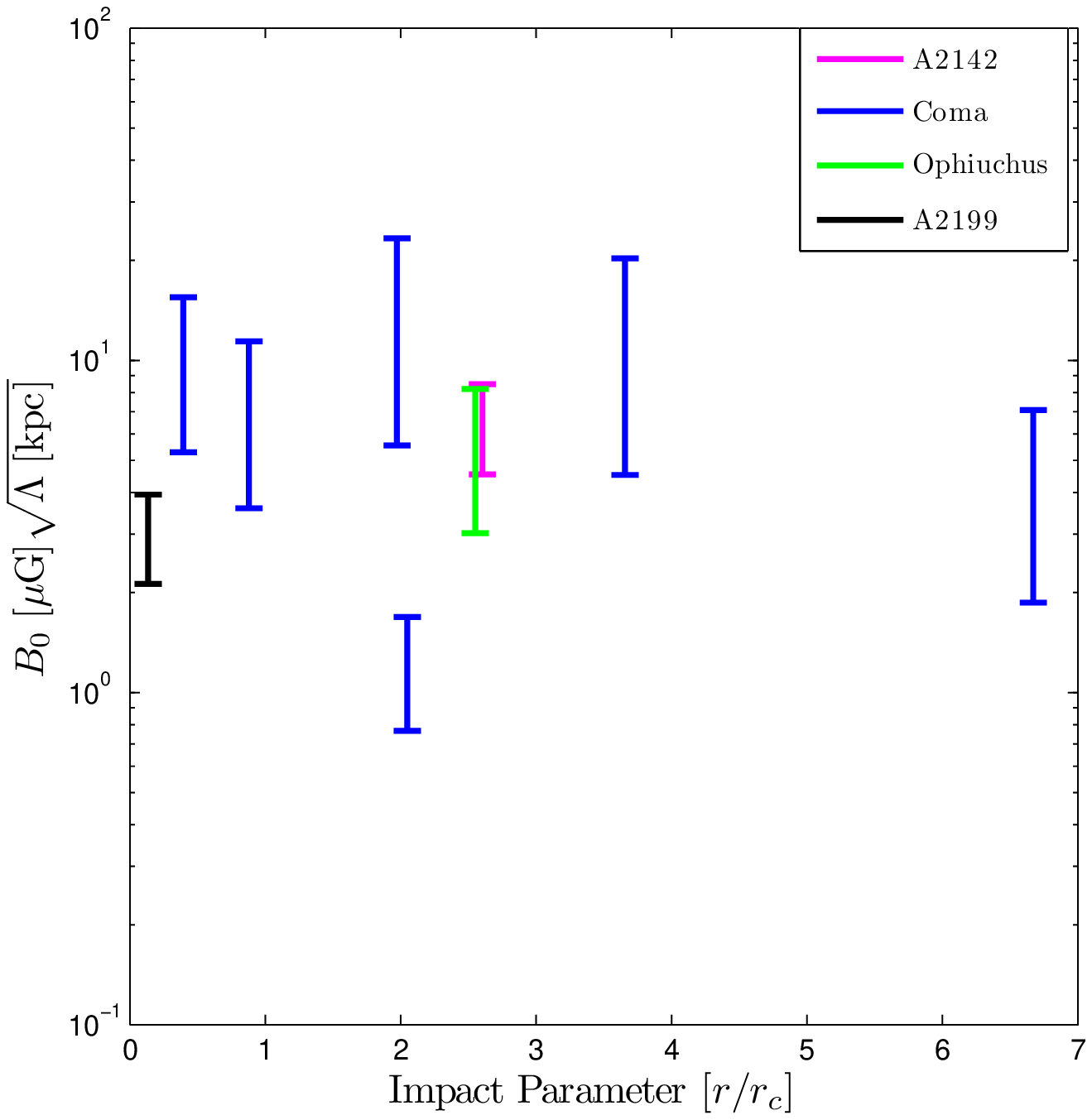} &
\includegraphics[width=0.48\linewidth]{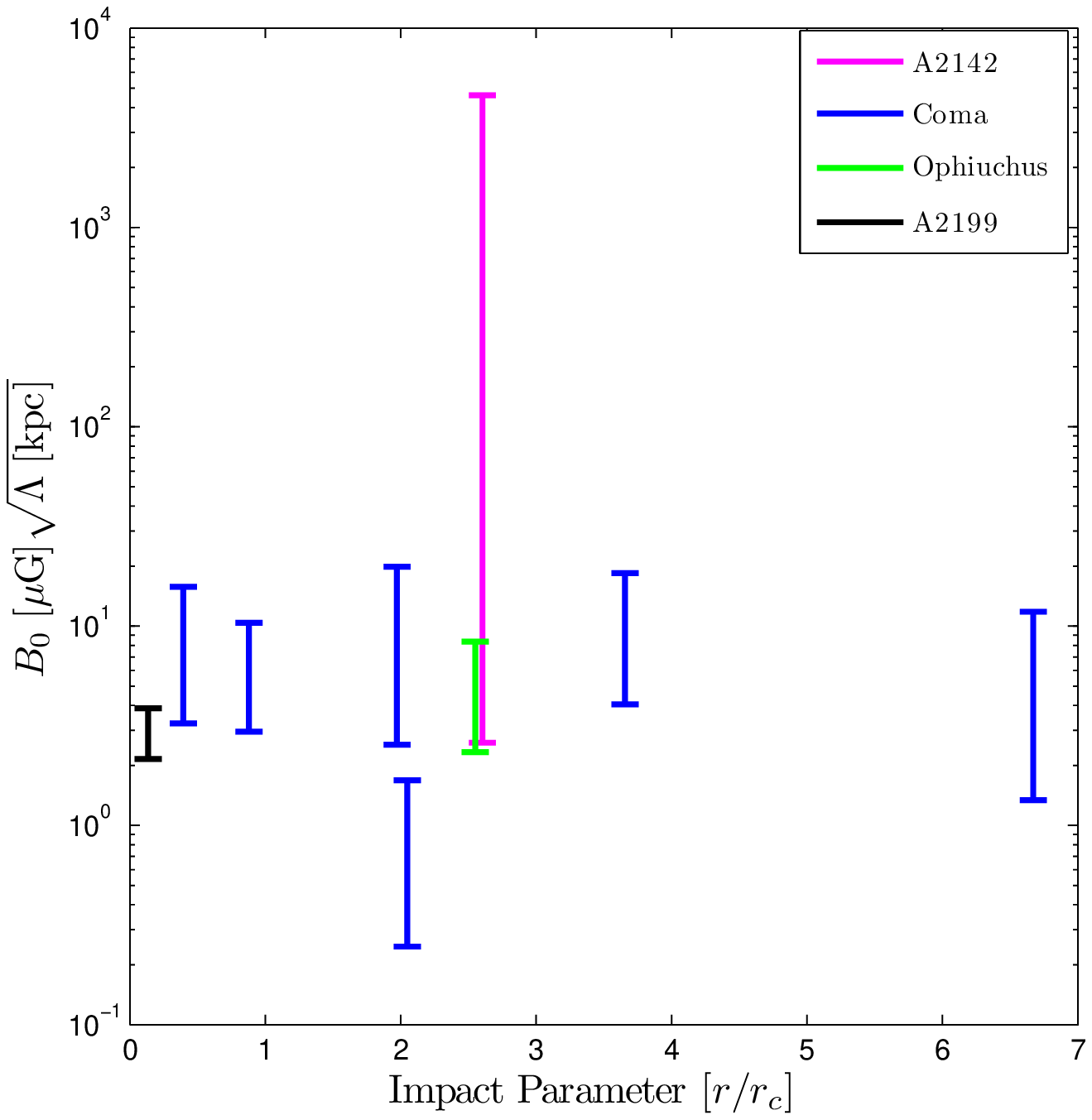} \\
\end{tabular}
\caption{The accepted ranges of magnetic field strengths of the varying component, $\Bc=(3\mean{B_z^2})^{1/2}$, for single-scale models. The top-left and top-right panels depict the accepted ranges of magnetic field strengths as a function of the impact parameter for the single-scale models without and with a homogeneous component respectively. Note that the y-axis scale differs between panels. In the bottom-left and bottom-right panels we show the corresponding $\Bc\Lambda^{1/2}$ values. The error bars represent the range of accepted values (95\% c.l.), and the arrows indicate the lower-bounds as derived in eq.~\ref{Eq:MinB0}.\label{Fig:Sng}}
\end{figure*}

For most sources, a single-scale model is consistent with the data. All accepted ranges of field strengths of our single-scale models, with and without a homogeneous component, are plotted in Figure~\ref{Fig:Sng} as a function of the projected distance from the cluster center. We also indicate
\begin{equation}
B\CMB \equiv \prnd{ 8 \pi a T\CMB^4 } ^{1/2} \simeq 3 \muG
\end{equation}
as a dashed gray line, where relevant, and notice that accepted field strengths are distributed around this line.

\begin{table}[!ht]
{\centering
\caption{Simulation results -- accepted models\label{Tbl:Results}}
\begin{tabular}{ccccc}
\toprule
	\ColHead{Tbl:Results:Src}	{Source}	{}			&
	\ColHead{Tbl:Results:Comp}	{Comp.}		{}			&
	\ColHead{Tbl:Results:BU}	{$\Bmsk$}	{\!\muG}	&
	\ColHead{Tbl:Results:B0}	{$\Bc$}		{\!\muG}	&
	\ColHead{Tbl:Results:L}		{$\Lambda$}	{\!\kpc}	\\
\midrule
	5C4.042	& U+P		& $\range{0.005}{0.025}$	& $\range{3.3}{7.6}$	& ---\tnt{Tbl:Results:Note:Mask}	\\
	5C4.074	& S / P		& $\le 0.57$				& $\range{0.3}{19}$		& $\range{1.3}{32}$	\\
	5C4.081	& S / P		& $\le 0.28$				& $\range{0.52}{5.9}$	& $\range{2}{32}$	\\
	5C4.085	& S / P		& $\le 0.46$				& $\range{0.52}{7.7}$	& $\range{3.3}{32}$	\\
	5C4.114	& U / S / P	& $\le 0.085$				& $\le 1.7$				& $\range{1}{32}$	\\
	5C4.127	& U+S / P	& $\le 0.41$				& $\range{1.0}{19}$		& $\range{1.5}{32}$	\\
	5C4.152	& S / P		& $\le 1.3$					& $\range{0.17}{16}$	& $\range{1}{32}$	\\
	A2065A	& \multicolumn{4}{c}{no models were accepted}										\\
	A2142A	& U / S / P	& $\le 0.3$					& $\range{0.2}{3700}$	& $\range{3.5}{32}$	\\
	3C 338	& S / P		& $\le 0.23$				& $\range{0.45}{3.9}$	& $\range{11}{28}$	\\
	OPHIB	& S / P		& $\le 0.19$				& $\range{0.5}{24}$		& $\range{4.9}{32}$	\\
\bottomrule
\end{tabular}
\\}
{\footnotesize
\begin{enumerate}
	\labitem{\dag}{Tbl:Results:Note:Mask}	Acceptable power-law indices for the source 5C4.042 range from $n=0$ to $n=2.4$.
\end{enumerate}
\textsc{Tab}. \ref{Tbl:Results}.---
\begin{inparaenum}[(1)]
	\item \label{Tbl:Results:Src}	Source name;
	\item \label{Tbl:Results:Comp}	Acceptable models ($95\%$ c.l.): U=uniform, S=single-scale, P=power-law, '/'=or, '+'=and.
									Note that a uniform component may always be added to a varying one;
	\item \label{Tbl:Results:BU}	Uniform component magnetic field strength;
	\item \label{Tbl:Results:B0}	Average magnitude of the spatially varying magnetic field component;
	\item \label{Tbl:Results:L}		Allowed spatial variation scales of the single-scale model.
\end{inparaenum}
}\\
\end{table}

\begin{figure*}[!t]
\centering
\begin{tabular}{cc}
\includegraphics[width=0.48\linewidth]{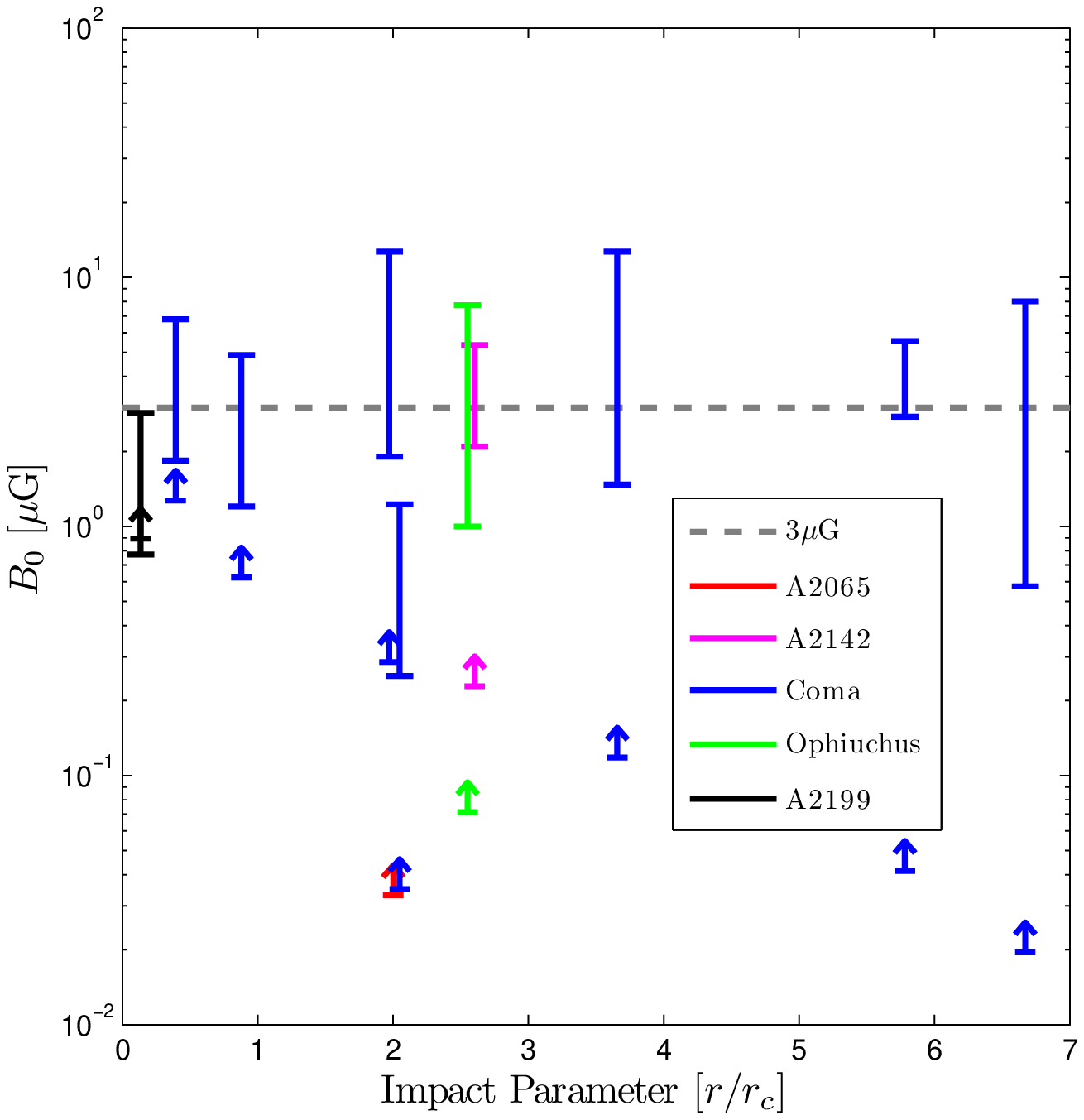} &
\includegraphics[width=0.48\linewidth]{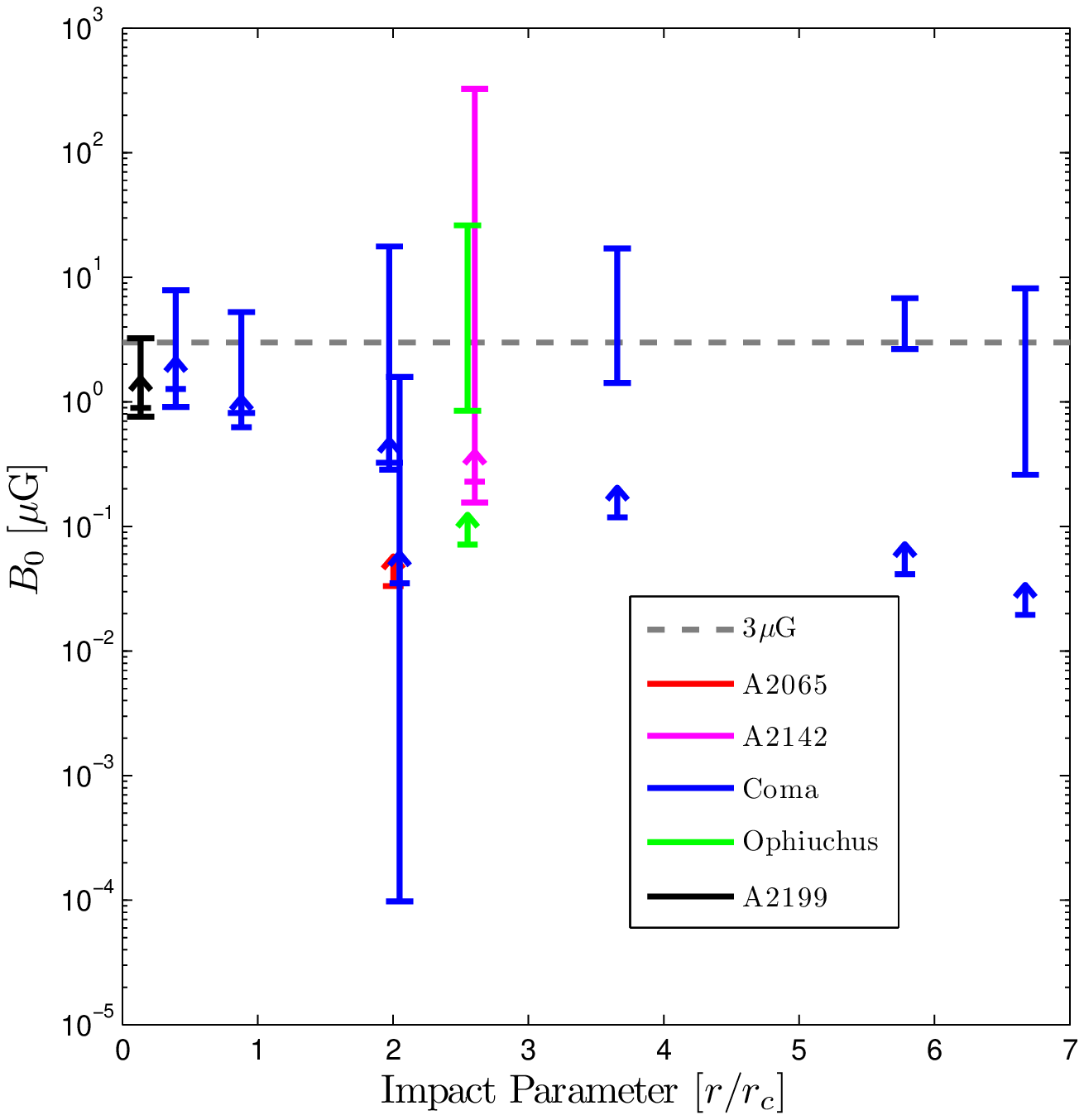} \\
\end{tabular}
\caption{The accepted ranges of magnetic field strengths of the varying component, $\Bc=(3\mean{B_z^2})^{1/2}$, for power-law models. The left and right panels depict the accepted ranges of magnetic field strengths for power-law models without and with a homogeneous component, respectively. Note that the y-axis scale differs between the panels. The error bars represent the ranges of accepted values (95\% c.l.), and the arrows indicate the lower-bounds as derived in Eq.~(\ref{Eq:MinB0}).\label{Fig:Pwr}}
\end{figure*}

All the radio sources analyzed in this work, except A2065A (the only source for which no model was accepted and may thus require a more complicated model), are consistent with some power-law magnetic field in the range of scales of $\Lmin=1\kpc$ to $\Lmax=32\kpc$, chosen to accommodate scales down to the angular resolution and up to the typical source size. In Figure~\ref{Fig:Pwr} we show the ranges of acceptable magnetic field strengths, for our power-spectrum models, along with the analytical lower bounds found using Eq.~(\ref{Eq:MinB0}). Again, there is no apparent radial trend of magnetic field strength, and the values are scattered around the $B\CMB$ line.

We note that although there is a correlation between the spectral index of the (3D) magnetic field power-spectrum and the resulting shape of the (2D) RM power-spectrum (e.g. the average RM power spectrum of a single-scale model peaks at the same scale as that of the magnetic field power spectrum), the spectral index of the magnetic field power spectrum is not given by a power-law fit to the RM power spectrum.


\section{Discussion} \label{Sec:Discussion}

We presented a simple and robust approach for deriving constraints on magnetic fields in galaxy clusters from rotation measure (RM) maps (\S~\ref{Sec:Methods}). Relaxing the commonly used assumptions of a correlation between the magnetic field strength and the plasma density \citep[e.g.][]{Bonafede2010,Vacca2012} and of a power-law (in wave number) magnetic field power spectrum, and using an efficient numerical analysis method, we tested the consistency of a wide range of magnetic field models with RM maps of 11 extended sources in 5 clusters (Coma, A2065, A2142, A2199, and Ophiuchus), for which the data were made available to us (\S~\ref{Sec:Results}). This constitutes a significant extension of previous work (\S~\ref{Sec:PrevWork}).

We have shown that the data reveal no indication for a radial dependence of the average magnetic field strength (see fig.~\ref{Fig:Sng}), and in particular no indication for a correlation between the gas density and the field strength. The RM maps of a considerable fraction of the sources either require or are consistent with the presence of a spatially uniform magnetic field of a relatively small strength (see table~\ref{Tbl:Results}), $0.02\text{--}0.3\muG$, which contributes significantly to the RM. The presence of such spatially uniform fields has already been suggested by \cite{Taylor1993} in their analysis of Hydra~A, but was not taken into account in recent studies \cite[][infer field values which are $\sim$10 times larger than our estimate of the uniform components in Coma, but the RM values in Hydra~A are also an order of magnitude larger]{Taylor1993}.

The RM maps of all but one source do not require a power-law magnetic field power spectrum, and most are consistent with a power spectrum dominated by a single wave length (see table~\ref{Tbl:Results}). The magnetic field values inferred by earlier analyses (\S~\ref{Sec:PrevWork}) are consistent with our results. However, we find that the uncertainties in the magnetic field strengths (and spatial correlation lengths) derived from RM maps are significantly larger than those inferred by earlier analyses, which considered a more limited range of models and model parameters. We find that the uncertainties exceed an order of magnitude and often more (see table~\ref{Tbl:Results}). This implies, in particular, that there is no indication in current RM data for a systematic difference between the magnetic field strengths in radio-halo clusters and in radio-quiet clusters. We note that the spatial characteristics of the 2D RM maps are highly affected by the beam resolution and by the shape of the underlying extended sources.


Our results have important implications also to the discussion of the mechanism responsible for the observed correlation \citep{Brunetti2007} between the radio power and the thermal X-ray luminosity of radio emitting galaxy clusters, which is a subject of considerable debate. Several models for the diffuse synchrotron emission, the so called radio halos (RHs), have been suggested, with different assumptions regarding the origin of the emitting electrons. In some models the emitting electrons (and positrons) are secondaries generated by p-p interactions between the CR protons and thermal ICM protons \citep[e.g.][]{Dennison1980, Kushnir2009b, Keshet2010}, whereas other models assume that the emitting electrons are re-accelerated by turbulence from a preexisting population of non-thermal seed electrons in the ICM \citep[secondary or otherwise, e.g.][]{Brunetti2001, Petrosian2001}. As pointed out by \cite{Kushnir2009b}, the upper envelope of the correlation between radio-power and X-ray luminosity is naturally obtained in models where the synchrotron radiation is produced by secondary electrons, provided the magnetic field strength exceeds $B\CMB=3\mu$G. In such models, the suppression observed in radio quiet galaxy clusters of the synchrotron luminosity, by a factor of $\sim10$ compared to the upper envelope of the correlation, must be due to lower values of the magnetic field, $B \lesssim 1\muG$. Therefore, accurate measurements of the magnetic fields in radio emitting and radio quiet galaxy clusters may discriminate between models \citep[see, however, ][for lack of such bi-modality between RH and radio quited clusters in Planck data]{Basu2012}.

The range of acceptable magnetic field strengths derived in this work implies that it is difficult to test the hypothesis of magnetic field bi-modality using RM maps (see the scatter around $B=B\CMB$ in figs.~\ref{Fig:Sng} \& \ref{Fig:Pwr}). Note, in particular, that we have shown here that there is no evidence to support some of the assumptions used in earlier works \citep[discussed in \S~\ref{Sec:PrevWork}, e.g.,][]{Bonafede2011,Govoni2010}, which argued against the possibility of such magnetic field strength bi-modality.

A further comment is in place here regarding the possible magnetic field bimodality. Based on a statistical analysis of the radial trend of depolarization, \cite{Bonafede2011} find no evidence for a difference between populations of galaxy clusters with and without a RH. They find, using the logrank test, that the null hypothesis that the values of depolarization for both populations (with and without a RH) are different realizations of the same distribution is acceptable. However, one should note that their analysis is based on RM measurements of sources that lie at distances from the cluster center that extend to $\sim10r_c$, while radio halos typically extend only up to $\sim2r_c$. It is unclear what the conclusion of the analysis would be if only sources up to this shorter distance are considered.

Accurate measurements of the magnetic fields in galaxy clusters might be possible in the near future, when the quality and quantity of RM data is improved. Higher sensitivity, resolution, and imaging capability will be provided by the EVLA extension of the VLA. Most important for RM maps, the resolution will be improved to $\range{0.02}{0.2}\unit{arcseconds}$ in the cm wavelength range, allowing one to explore the contribution of shorter spatial fluctuation scales. LOFAR will explore the universe at radio-frequencies below $250\,\textrm{MHz}$, and will thus be able to detect diffuse cluster radio sources which are brighter at lower frequencies, in particular radio halos with very steep spectra \citep{Feretti2012}. The longer wavelengths also make it possible to detect lower RM values, allowing one to fill some of the missing areas in current RM maps.

New wide-field surveys of the sky will offer one the opportunity to carry out RM studies of many more galaxy clusters.
Two such projects are planned with the ASKAP telescope. EMU is a radio sky survey project which will make a deep ($\sim10\,\mu\textrm{Jy}$ r.m.s.) radio continuum survey covering the entire Southern Sky. It will have about 45 times the sensitivity of the NVSS, and an angular resolution 4.5 times better \citep{Cassano2012}. Because of the excellent short-spacing UV coverage of ASKAP, EMU will also have higher sensitivity to extended structures such as cluster halos \citep{Cassano2012}. In POSSUM, the plan is to use ASKAP's unique survey capabilities to measure the Faraday rotation of three million extragalactic radio sources over 30,000 square degrees ("POSSUM Wide").

APERTIF, the new Phased Array Feed system that will be installed on WSRT, will increase by about a factor 30 the observed area on the sky, at frequencies of $1.0\GHz$ to $1.7\GHz$ \citep{Cassano2012,Rottgering2011}. The survey speed of APERTIF, and many of the other characteristics, will be very similar to ASKAP. The extremely large field of view of APERTIF would enable the WODAN project, which has been proposed with the aim of charting the entire accessible northern sky at $1.4\GHz$ down to $10\,\mu\textrm{Jy}$ r.m.s. and $\sim1000\unit{deg}^2$ down to $5\,\mu\textrm{Jy}$ \citep{Cassano2012}.

Other instruments, such as the LWA, MeerKAT, and the African SKA, will also contribute to increasing the quantity and quality of RM observations.

As RM data becomes more abundant, with more sources being observed, closer to and within galaxy cluster cores, our proposed analysis method will provide higher quality estimates of the magnetic field strength. Furthermore, it may prove useful to exploit the extra information encoded in the polarization maps themselves (from which the RM maps are derived), as has recently been done for A2199 \citep{Vacca2012}, in order to better constrain magnetic field models.


\acknowledgments

The authors thank F. Govoni, A. Bonafede, V. Vacca and their collaborators for the contribution of the RM data and useful discussions. D.~K is supported by NSF grant AST-0807444. GR \& EW are partially supported by a UPBC grant.


\appendix


\section{From ICM Magnetic Fields To RM Maps}\label{Apx:B2RM}

We derive below an analytic expression for the RM, eq.~(\ref{Eq:RM}), obtained for a magnetic field of the form given by eq.~\eqref{Eq:Bk},
\begin{align}
\Br	
	= N \Dk \Du \Df \Bc \sum_{k=\kmin}^{\kmax} k^{2-\frac{n}{2}}  \mathbf{F}_{k},
\end{align}
with
\begin{equation}\label{eq:Fk}
   \mathbf{F}_{k} =\sum_{u=-1}^{+1} \sum_{\phi=-\pi}^{+\pi} e^{i \bOm}
		\psqr{ \khat_1 \cos\of{\aOm}
			+ \khat_2 \sin\of{\aOm}} e^{i \kdotr}.
\end{equation}
Here, $\Df,\Du,\Dk$ are the $k$-space resolution of the azimuthal, (cosine of the) polar, and radial (spherical) coordinates, respectively, and $N$ is a normalization factor to be determined below by requiring $\Bc=(3\mean{B_z^2})^{1/2}$. Writing
\begin{align}
\mathbf{F}_{k} = \sum_{u=0}^{+1} \sum_{\phi=-\pi}^{+\pi} & \psqr{
				e^{i \bOm} \prnd{ \khat_1 \cos\of{\aOm}
										+ \khat_2 \sin\of{\aOm} } e^{+ i \kdotr} \right.
		\nonumber	\\	&\quad \quad \left. + e^{i \beta_{-\kvec}} \prnd{ - \khat_1 \cos\of{\alpha_{-\kvec}}
					- \khat_2 \sin\of{\alpha_{-\kvec}} } e^{- i \kdotr} }
\end{align}
and choosing $\alpha_{-\hat{\Omega}}=\aOm$ and $\beta_{-\hat{\Omega}}+\bOm=\pi$, so that $\Br$ is real, we have
\begin{align}
\mathbf{F}_{k} &= \sum_{u=0}^{+1} \sum_{\phi=-\pi}^{+\pi}
				\prnd{\khat_{1}\cos\of{\aOm}+\khat_{2}\sin\of{\aOm}}
				\psqr{e^{+i\prnd{\bOm+\kdotr}}+e^{-i\prnd{\bOm+\kdotr}}}\nonumber
\\ &= 2 \sum_{u=0}^{+1} \sum_{\phi=-\pi}^{+\pi}
				\prnd{\khat_{1}\cos\of{\aOm} + \khat_{2}\sin\of{\aOm}} \cos\of{\bOm+\kdotr}
\end{align}
and
\begin{align}
B_z\of{\rvec} = 2 N \Bc \sum_{\kvec} \cos\of{\aOm} \cos\of{\bOm + \kdotr} k^{2-\frac{n}{2}} \Dk\Du\Df  \sqrt{1-u^2},
\end{align}
where the factor $\sqrt{1-u^2}$ arises from the $z$ component of $\khat_1$ (the $z$-component of $\khat_2$ vanishes).
The average magnetic field strength is then given by
\begin{align}
\mean{\prnd{B_z\of{\rvec}}^2} &= 4 N^2 \Bc^2 \sum_{\kvec_1,\kvec_2}
		 \underbrace{\mean{\cos\of{{\aOm}_{_1}}\cos\of{{\aOm}_{_2}}}}
					_{=\frac{1}{2}\delta_{\kvec_1,\kvec_2}}
		\underbrace{\mean{\cos\of{{\bOm}_{_1}+\kvec_{1}\cdot\rvec}
								\cos\of{{\bOm}_{_2}+\kvec_{2}\cdot\rvec}}}
					_{=\frac{1}{2}\delta_{\kvec_1,\kvec_2}}\nonumber \\ \nonumber
	&	\hspace{1.0in} \prod_{i=1,2} k_i^{2-\frac{n}{2}} \Dk_i\Du_i\Df_i  \sqrt{1-u_i^2} \\
	&=  N^2 \Bc ^2 \sum_{k,u,\phi} k^{4-n} \prnd{\Dk\Du\Df}^2 \prnd{1-u^2},
\end{align}
which is independent on $\rvec$. Defining $N^2=N'/\left(\Dk\Du\Df\right)$,
\begin{align}
\mean{B_z^2}
	&= N' \Bc^2 \sum_{k,u,\phi} k^{4-n} \Dk\Du\Df \prnd{1-u^2}
\nonumber\\	&\approx N' \Bc^2 \int_{\kmin}^{\kmax} k^{4-n} \drm k
		\int_{0}^{1} \prnd{1 - u^2} \drm u \int_{0}^{2\pi} \drm \phi
\nonumber\\	&= N' \Bc^2  \frac{\kmax^{5-n} - \kmin^{5-n}}{5-n} \frac{2}{3} 2\pi,
\end{align}
we find
\begin{align}
N' = \frac{1}{4\pi} \frac{5-n}{\kmax^{5-n} - \kmin^{5-n}},
\end{align}
and the final expression for the magnetic field is
\begin{align}
B_z\of{\rvec} = \Bc\frac{1}{\sqrt{\pi}}\sum_{k=\kmin}^{\kmax}\sum_{u=0}^{1}\sum_{\phi=0}^{2\pi}
		N_k N_{\Omega} \sqrt{1-u^2} \cos\of{\aOm} \cos\of{\bOm+\kdotr}.
\end{align}
Here
\begin{align}
N_k^2 &= \frac{\prnd{5-n}k^{4-n}\Dk}{\kmax^{5-n}-\kmin^{5-n}}
	\hspace{0.1in} , \hspace{0.2in} N_{\Omega}^2 = \Du\Df.
\end{align}
The rotation measure, $\RM$, is now given by
\begin{align}
\RM\of{x,y} &= \RM_0 \int B_z n_e \drm z \\ \nonumber
	&= \RM_0 \Bc n_0 r_c \sum_{k,u,\phi} N_k N_{\Omega}
			\sqrt{\prnd{1-u^2}} \cos\of{\aOm} \\
	& \hspace{0.5in} \times \sqrt{2}\underbrace{\frac{1}{\sqrt{2\pi}r_c} \int \drm z \cos\of{\bOm+\chi+kuz}
						\prnd{1+\frac{x^2+y^2+z^2}{r_c^2}}^{-b}}_{\equiv I},
\end{align}
where
\begin{equation}
b = \frac{3}{2}\beta
	\hspace{0.1in} , \hspace{0.2in} \chi=k\sqrt{1-u^2}\prnd{x\cos\phi+y\sin\phi}.
\end{equation}
One may write
\begin{equation}
I = I_0 \Re \psqr{ e^{i\prnd{\beta_{\hat{\Omega}}+\chi}}
				\frac{1}{\sqrt{2\pi}}\int\prnd{1+\zeta^2}^{-b}e^{i\omega\zeta}\drm\zeta },
\end{equation}
where
\begin{equation}
I_0 = \prnd{ r / r_c } ^{1-3\beta}
	\hspace{0.1in} , \hspace{0.2in} \zeta = z / r
	\hspace{0.1in} , \hspace{0.2in} \omega = k u r
	\hspace{0.1in} , \hspace{0.2in} r = \sqrt{r_c^2+x^2+y^2} \hspace{0.1in} .
\end{equation}
For a background source, the integral over the line of sight ($\int_{-\infty}^{+\infty}$) is a Fourier transform,
\begin{equation}
\mathcal{F}_{\zeta}\pcrl{\prnd{1+\zeta^2}^{-b}}\of{\omega}
	= \frac{2^{1-b}}{\Gamma\of{b}}
			\abs{\omega}^{b-\frac{1}{2}}K_{b-\frac{1}{2}}\of{\abs{\omega}},
\end{equation}
where $K$ is the modified Bessel function of the second kind, and the final expression for the rotation measure is
\begin{align}
\RM\of{x,y} &=  \RM_0 \Bc n_0 r_c
			\sum_{k,u,\phi} N_k N_{\Omega} \frac{2\sqrt{1-u^2}}{\Gamma\of{\frac{3}{2}\beta}}
			\prnd{\frac{ku r_c^2}{2r}}^{\frac{3}{2}\beta-\frac{1}{2}}
			K_{\frac{3}{2}\beta-\frac{1}{2}}\of{kur} \nonumber\\
	& \hspace{0.5in} \times\cos\of{\alpha_{\hat{\Omega}}}
						\cos\of{\beta_{\hat{\Omega}}+k \sqrt{1-u^2}\prnd{x\cos\phi+y\sin\phi}}.
\end{align}
For a cluster member the integral over the line of sight ($\int_{0}^{\infty}$) is a Fourier Cosine/Sine transform,
\begin{align}
\frac{1}{2} \psqr{\mathcal{FC}+i\mathcal{FS}}_{\zeta}
				& \!\pcrl{\prnd{1+\zeta^2}^{-b}}\of{\omega} \nonumber\\
	& = \frac{1}{2}\frac{2^{1-b}}{\Gamma\of{b}}\omega^{b-\frac{1}{2}}\psqr{ K_{b-\frac{1}{2}}\of{\omega}
		+i\pi\frac{I_{b-\frac{1}{2}}\of{\omega}-L_{\frac{1}{2}-b}\of{\omega}}{2\sin\of{b\pi}} },
\end{align}
where $I$ is the modified Bessel function of the first kind, and $L$ is the modified Struve function, and the final expression for the rotation measure of a cluster member is
\begin{align}
\RM\of{x,y} &=  \RM_0 \Bc n_0 r_c \sum_{k,u,\phi} N_k N_{\Omega}
						\frac{\sqrt{1-u^2}}{\Gamma\of{\frac{3}{2}\beta}}
						\prnd{\frac{ku r_c^2}{2r}}^{\frac{3}{2}\beta-\frac{1}{2}}
						\cos\of{\alpha_{\hat{\Omega}}}\nonumber \\
	& \times \pcrl{\cos\of{\beta_{\hat{\Omega}}+\chi} K_{\frac{3}{2}\beta-\frac{1}{2}}\of{kur}
		+ \sin\of{\beta_{\hat{\Omega}}+\chi}
					\pi \frac{ L_{\frac{1}{2}-\frac{3}{2}\beta}\of{kur}
							- I_{\frac{3}{2}\beta-\frac{1}{2}}\of{kur} }
						{2\sin\of{\frac{3}{2}\beta\pi} } }.
\end{align}


\bibliographystyle{hapj}
\bibliography{FRM}

\end{document}

%% file: Packages.tex
\usepackage{amsmath}	
\usepackage{amssymb}	%
\usepackage{bm}			
\usepackage{booktabs}	
\usepackage{color}		
	\definecolor{gray}{rgb}{0.5,0.5,0.5}	
	\definecolor{dred}{rgb}{0.5,0,0}		
	\definecolor{dgreen}{rgb}{0,0.5,0}		
	\definecolor{dblue}{rgb}{0,0,0.5}		
	\definecolor{orange}{rgb}{1.0,0.5,0.0}	
\usepackage{graphicx}	
\usepackage{epstopdf}	
\usepackage{hyperref}	
	\hypersetup{
		bookmarks=true,				
		citecolor=dgreen,			
		colorlinks=true,			
		filecolor=gray,				
		linkcolor=dred,				
		linktoc=section,			
		pdfauthor={Gilad Rave},		
		pdfkeywords={Magnetic Field}{Galaxy Cluster}{Faraday Rotation},
		pdfnewwindow=true,			
		pdfstartview={FitH},		
		pdfsubject={What can we really learn about Magnetic Fields in Galaxy Clusters from Faraday Rotation observations?},
		pdftitle={FRM},				
		urlcolor=blue,				
	}
\usepackage{natbib}		
	\bibpunct{(}{)}{;}{a}{}{,}
\usepackage{paralist}	
\usepackage[flushleft,online]{threeparttable}
\usepackage{ulem}		

%% file: Definitions.tex
\newcommand{\prnd}[1]{\left(#1\right)}
\newcommand{\psqr}[1]{\left[#1\right]}
\newcommand{\pcrl}[1]{\left\{#1\right\}}
\newcommand{\mean}[1]{\left\langle#1\right\rangle}
\newcommand{\abs}[1]{\left|#1\right|}

\newcommand{\of}[1]{\!\left(#1\right)}

\newcommand{\dfracn}[3]{\frac{\drm^{#2}#3}{\drm{#1}^{#2}}}

\DeclareBoldMathCommand{\Nabla}{\nabla}

\newcommand{\drm}{\mathrm{d}}
\newcommand{\unit}[1]{\, \mathrm{#1}}
\newcommand{\kpc}{\unit{kpc}}
\newcommand{\muG}{\unit{\mu G}}
\newcommand{\GHz}{\unit{GHz}}

\newcommand{\Mpc}{\unit{Mpc}}
\newcommand{\s}{\unit{s}}
\newcommand{\rad}{\unit{rad}}
\newcommand{\m}{\unit{m}}
\newcommand{\cm}{\unit{cm}}
\newcommand{\km}{\unit{km}}

\newcommand{\Msun}{M_{\odot}}

\newcommand{\Bc}{B_0}
\newcommand{\Bmsk}{B_{\text{hom}}}
\newcommand{\kmax}{k_{\text{max}}}
\newcommand{\kmin}{k_{\text{min}}}
\newcommand{\Lmax}{\Lambda_{\text{max}}}
\newcommand{\Lmin}{\Lambda_{\text{min}}}
\newcommand{\aOm}{\alpha_{\hat{\Omega}}}

\newcommand{\bOm}{\beta_{\hat{\Omega}}}

\newcommand{\Dk}{{\Delta k}}
\newcommand{\Du}{{\Delta u}}
\newcommand{\Df}{{\Delta \phi}}
\newcommand{\kvec}{\mathbf{k}}
\newcommand{\Bvec}{\mathbf{B}}
\newcommand{\Br}{\Bvec\of{\rvec}}

\newcommand{\RM}{\mathrm{RM}}

\newcommand{\sRM}{\sigma_{\RM}}
\newcommand{\khat}{\hat{\kvec}}
\newcommand{\rvec}{\mathbf{r}}
\newcommand{\kdotr}{\kvec\cdot\rvec}
\newcommand{\CMB}{_{\text{CMB}}}

\newcommand{\range}[2]{#1\text{--}#2}

\newcommand{\txtsup}[1]{\textsuperscript{#1}}
\newcommand{\tnt}[1]{\txtsup{\ref{#1}}}
\newcommand{\ColHead}[3]{\begin{tabular}[t]{c}	(\ref{#1})\\
												#2\\
												\if x#3x\else$[#3]$\\\fi	\end{tabular}}

\makeatletter
\newcommand{\labitem}[2]{%
\def\@itemlabel{\textbf{#1}}
\item
\def\@currentlabel{#1}\label{#2}}
\makeatother
\makeatletter
	\renewcommand\p@enumii{} 
\makeatother